\documentclass{elsart}

\usepackage{graphics}
\usepackage{amssymb}

\begin{document}

\begin{frontmatter}



\title{The approach to steady state in microemulsions under shear flow}


\author{Domenico Suppa}
\ead{suppa@thphys.ox.ac.uk}
\address{Theoretical Physics, University of Oxford,\\
1 Keble Road, OX1 3NP, U.K.}

\begin{abstract}
We present an analitical study of the dynamical process
of the approach to steady state for a driven diffusive system 
represented by the microemulsion phase of a ternary mixture. The 
external applied field is given by a \textquotedblleft plane 
Couette\textquotedblright  shear flow and the problem is studied 
within the framework of a time-dependent Ginzburg-Landau model. 
A Fokker-Planck equation for the probability distribution of the 
concentration fluctuations is developed in a self-consistent one-loop
approximation and solved exactly, giving an analytical expression for
the dynamic correlation functions of the system. For comparison to
experimental work, we also show grey-scale plots
of the scattering function at different times during the 
dynamical process for two different shear rates.
\end{abstract}

\end{frontmatter}

% main text
\section{Introduction}
\label{Introduction}
Ternary mixtures made up of oil, water and surfactant agent are of
considerable interest for both fundamental issues as well as for
applications \cite{Meu,Mic,Gom1}. In these systems, three-phase
coexistence is commonly observed between a water-rich phase, an
oil-rich phase and a microemulsion. The latter is a mesoscopic structured
phase where coherent domains of oil and water on scales between 100 
and 1000 $\AA$ form disordered isotropic intertwined structures on 
larger scales \cite{DeG}. A very similar mesoscopic structure is
observed in sponge phases in acqueous amphiphilic solutions
\cite{Por,Rou} and in homopolymer blends containing small amounts
of diblock copolymers which act as a surfactant \cite{Bin,Kos,Buck,Wag,Ada}.

After the earlier work by Gompper and Shick \cite{Gom2} in which they
demonstrate that the failure of the microemulsion to wet the oil-water
interface is a consequence of the microemulsion internal structure,
various theoretical advances on similar ternary systems have been
made \cite{Gom3,Wid,Jan,Mut2}; in particular Kielhorn and Muthukumar
\cite{Mut2} have investigated the weak segregation behaviour of a
ternary mixture of A homopolymer, B homopolymer and AB diblock
copolymer, showing several features which are identical to those seen
in traditional oil/water/surfactant systems.
These amphiphilic systems are now receiving renewed attention, since
several light and neutron scattering experiments revealed the
hallmarks of glassy dynamics, in the sense of large relaxation times, on 
their mesoscopic structures \cite{Choy,Gang,Sheu,Sollich,Struik,Bouch}.
Other ways of establishing non-equilibrium states with slow
dynamics are also possible. For instance,
microemulsions can be driven to a slow out of equilibrium
stationary regime by a weak shear. This perturbation accelerates the
dynamics in such a way that the structural relaxation time 
decreases with increasing shear strength (a phenomenon usually
known as shear-thinning). The shear flow can therefore be used as a probe
of the nonequilibrium properties of these systems by adopting the shear
rate $\dot{\gamma}$ as a control parameter \cite{Barrat}.
Several scattering experiments have shown the very complex and diverse 
effects of shear flow on these systems after the steady-state has been
reached \cite{Lar,WuPi,Roux,Boue,Yam,Rus1,Mig,Rus2,Bonn}: 
nonlinear effects due to the flow give rise to strongly 
anisotropic suppression of the composition fluctuations; the phase 
diagram is affected by the flow which can induce order-disorder 
transitions; the rheological behaviour is highly non-Newtonian with significant intriguing shear-thinning and shear-thickening behaviour.\\
In the following, we shall find useful to refer to the concentration fluctuation 
(equal time) correlation function, calling it ``structure factor'',
although such a name usually indicates the density fluctuation.

Theoretical efforts to understand these complex problems have also
been made intensively.
A mean field theory able to describe the scattering function of a
diblock copolymer in the presence of a steady homogeneous flow field
was developed by Fredrickson \cite{Fred1} as a generalization of the
equilibrium theory of Leibler \cite{Lei}. The theory predicts the
wave vector dependence of the scattering intensity and the distribution
function of composition fluctuations in the disordered phase.
The concept of an increase of the order-disorder transition temperature
due to the shear was first studied for small shear rates by Cates and Milner 
\cite{CatMil} who applied a self-consistent Hartree approximation to 
the nonequilibrated system and solved for the shear effect on this Hartree term. A more general self-consistent approach, not limited to small shear rates 
and with predictions which are in good agreement with experimental 
results \cite{Kop,Bal,Nak}, was developed by Huang and 
Muthukumar \cite{Mut1} who studied numerically the effect of shear 
rate on order-disorder and order-order transitions 
involving lamellar and cylindrical morphologies as well as the dependence 
of steady-state scattering function on shear.
A one-loop self consistent Ginzburg-Landau model for describing the 
steady-state rheological behavior of  oil/water/surfactant mixtures
under a plane Couette shear flow was considered by P\"atzold and 
Dawson \cite{Daw1} who gave a numerical evaluation of the structure
factor together with an analytical analysis in the limit of vanishing 
shear rate $\dot{\gamma}$.
Their model was exactly solved in \cite{Sup1} for any value of 
$\dot{\gamma}$ showing that in a specific shear range
the structure factor is characterized by four pronounced peaks in 
the plane of the shear and flow directions. This phenomenon, which 
was already predicted by Drolet et al. in a slightly different model
\cite{Dro} and also reported in scattering experiments on segregating
mixtures \cite{Mig,WuPi}, was interpreted in terms of a complex spatial pattern where interfaces with two different orientations coexist, with relative 
abundance tuned by $\dot{\gamma}$.
On the other hand, the relaxation dynamics of microemulsions towards a steady
state, which can be extremely slow, is also very intriguing: in fact, the absence of time translation invariance gives rise to a complex behaviour where
different modes relax on different time scales and differently compete with the hydrodynamics time scales.

A systematic study of the dynamical behaviour of microemulsions
and sponge phases in thermal equilibrium and in the presence of
hydrodynamic modes has been developed by Hennes and Gompper in 
\cite{Gom4}: they use a field-theoretic perturbation theory which
makes use of \textquotedblleft response fields \textquotedblright 
\cite{Mar,Bau,DeDo} and whose key ingredient is a self-energy matrix
having a block structure identical to the Gaussian correlation 
and response matrix.
Within that framework, they were able to calculate the 
frequency-dependent (complex) viscosity, the sound velocity and 
damping  in a one-loop approximation.
However, the macroscopic relaxation of the system
and the role of the different time scales that are involved
are most clearly shown by the evolution of the scattering intensity and,
in particular, its asymptotic behaviour for short and large times.
This was calculated in \cite{Gom4} by using the first Born
approximation for evaluating the correlation matrix and,
as it was already pointed out by those authors, it is not
clear to what extent such behaviour relies  
on the adopted approximation. As a result, a self-consistent 
calculation of the dynamic behaviour of microemulsions and 
sponge phases is expected to clarify this important point.

Of concern in this paper is the microemulsion phase only and 
its slow relaxation dynamics towards the steady state, associated 
with composition fluctuations under a plane Couette shear flow.
We will focus on the time-dependent static structure factor (i.e. the
average of the concentration fluctuation at equal time, whose time dependence
comes from the transition from equilibrium to steady state), and perform
a self-consistent one-loop calculation within the model previously
adopted in \cite{Sup1}, which has the significant feature that results
can be formulated in terms of the physical correlation lengths, thereby
making direct comparisons to experiment possible. The aim of the present
work is to contribute in making a contact between the mesoscopic 
structural characteristic of microemulsions and
the general macroscopic dynamic features of soft materials.
The plan of this paper is as follows. In Sec.~2 we present the adopted
Ginzburg-Landau model for the microemulsion and derive the Fokker-Planck
equation for the probability distribution of composition fluctuations. In 
Sec.~3 we obtain its dynamic solution which gives an analytical
expression for the time-dependent static structure factor. We discuss the 
scattering function convergence towards its steady-state value
and show its morphologies during the relaxation process. 
In Sec.~4 we draw our conclusions.

\section{Ginzburg-Landau Model}
\label{Ginzburg-Landau Model}
We shall consider a binary fluid of oil and water at 50:50 concentration,
with the addition of non-ionic surfactant agent, 
in thermal equilibrium at a temperature $T>T_c$.
In order to describe the ternary mixture, we adopt a Ginzburg-Landau
model with a single scalar order parameter $\varphi$ representing
the local concentration difference between oil and water components.
The following coarse-grained Ginzburg-Landau free-energy, firstly proposed 
by Gompper and Schick \cite{Gom1}, has been used extensively for 
investigations of the static structure and phase behavior of these 
systems \cite{TeuStr,Gom1,Gom2}:
\begin{equation}
F[\varphi] = \int d^3 x \left( 
f(\varphi) + \frac{1}{2}\: g(\varphi) \mid \vec{\nabla} \varphi \mid^2
+\: \frac{c}{2}\: ( \nabla^2 \varphi )^2 \right)
\label{eqn1}
\end{equation}
where $f(\varphi)$ is a polynomial function of $\varphi$
which has three minima $\varphi_{1}\:<\:\varphi_{2}\:<\:\varphi_{3}$, with 
$\varphi_{2}\:=\:0$, corresponding to the three homogeneous bulk 
phases, oil-rich, microemulsion and water-rich respectively. Moreover
$g(\varphi)=g_0 + g_2 \varphi^2$, with $g_0<0$ and $g_2>0$, in order 
to grasp the experimental evidence that the peak in the structure
factor in the oil and water phases is at zero wavevector (so that 
$g(\varphi)$ must be sufficiently positive there), while the peak 
is at non-zero wavevector in the microemulsion (so that $g(\varphi)$ 
must be negative there).
The last term, with a coefficient $c>0$, makes stable the free-energy 
at large momenta and weights the curvature of the interfaces where surfactant 
gathers. 
As we are concerned with the microemulsion phase, \textit{i.e.}
the disordered but structured region of the phase diagram where $a>0$
and a stability condition $4 a c - g_0^2 > 0$ is fulfilled,
we will use a piecewise form of $f(\varphi)$:
\begin{equation}
f(\varphi)=\frac{a}{2} \varphi^2 + \frac{b}{4!} \varphi^4 
\label{eqn2}
\end{equation}
as proposed by P\"atzold and Dawson \cite{Daw1}, where the
quartic term has been included in order to capture the fluctuation effects
in the critical region.
Although local concentration of surfactant does not appear in
the free-energy, its average value is taken into account through
the coefficients $g_0$ and $c$: in particular the
negative value of $g_0$, known as the surface tension parameter, 
allows us to model a characteristic feature of the microemulsion phase,
\textit{i.e.} the tendency of surfactant to create spontaneously interfaces.
More refined models should include an additional order parameter 
to describe the density of surfactant (for a review about this point
see \cite{Gom1}).
In the following we will be concerned with the response 
of the system to a plane Couette shear flow where the average
velocity field is expressed as
\begin{equation}
\label{eqn_3}
\vec v(\vec x) = \dot{\gamma}\: y \: \vec e_{x}
\end{equation}
where $\dot{\gamma}$ is the shear rate and has the dimensions of a frequency.
The reference frame adopted is such that the x-axis is along the flux,
the y-axis is along the velocity gradient and the z-axis is along the
vorticity direction. $\vec e_{x}$ appearing in Eq.~(\ref{eqn_3}) is the 
unit vector in the flux direction. 
We will only consider shear flows such that the molecular relaxation 
times are less than the inverse of the shear rate $\dot{\gamma}$, 
\textit{i.e.} the Deborah number $D=\dot{\gamma}\tau <<1$. 
Moreover, we will assume that, due to the highly viscous nature of the 
mixture, the fluctuations of the velocity field are small and that their
coupling with the fluctuations of the order parameter is negligible
\cite{Fred1,Fred2}.
Under these assumptions it is legitimate to apply the stochastic
model given by Onuki and Kawasaki \cite{OK1} and study in the momentum 
space the composition fluctuations by the following Langevin-like equation
\begin{equation}
\label{eqn_4}
\frac {\partial}{\partial t} \varphi(\vec k, t) =
\dot{\gamma} \:k_x\: \frac{\partial}{\partial k_y} \varphi(\vec{k},t)
-\Gamma k^2  \left(\frac{\delta F }{\delta \varphi(- \vec k, t)}
\right) + \eta(\vec k, t)
\end{equation}
where the Onsager coefficient $\Gamma$ is assumed to be a constant 
at a temperature T sufficiently higher than $T_c$ (this assumtion
cannot be extended to homopolymer mixtures where
a k-dependent Onsager coefficient must be used \cite{Bin2}) and 
$\delta F / \delta \varphi$ is the 
usual thermodynamic force. $\eta$ is a stocastic white noise 
describing thermal fluctuations into the system: it is Gaussian
distributed with mean zero and variance constructed to satisfy a
(generalized) fluctuation-dissipation theorem \cite{Kubo}:
\begin{eqnarray}
\label{eqn_5a}
\langle \eta (\vec{k}, t) \rangle &=& 0 \\
\langle \eta (\vec{k}, t) \:\eta (\vec{k'}, t') 
\rangle &=& 2\:\Gamma \: k^2\: 
\delta(\vec{k}+\vec{k'}) \:\delta(t-t')
\label{eqn_5b}
\end{eqnarray}
As usual, it is assumed that the fluctuating force is uncorrelated
in time, which implies that one considers times much larger than the
characteristic times during which internal modes equilibrate.
Thereby, the stochastic process becomes Gaussian and Markovian 
\cite{Kub}.
In order to linearize the problem, we further assume that it is 
legitimate to approximate the term $\delta F[\varphi]\:
/\: \delta \varphi(- \vec k, t)$ as $\chi(k)\: \varphi(\vec k, t)$, 
where $\chi^{-1}(k)$ can be regarded as a static susceptibility that 
depends on the correlation functions evaluated in the steady state.
A self-consistent one-loop approximation gives
\begin{equation}
\label{eqn_6} 
\chi (k)= a^R + g^R k^2 + c k^4 
\end{equation}
with $a^R=a\:+\:b/2 \:S_0^\infty \:+\:g_2\: S_2^{\infty}$, 
 $g^R= g_0 \:+\: g_2\: S_0^{\infty}$, and  
\begin{equation}
S_n^{\infty} =\lim_{t \to \infty}\: \int _{|\vec k|<\Lambda}  \frac {d\vec k}
{(2\pi)^3} \mid \vec k \mid^n C(\vec k,t)
\label{eqn_7}
\end{equation}
In Eq.~(\ref{eqn_7}), $C(\vec k, t)= \langle \varphi(\vec k, t) \varphi(-\vec k, t) 
\rangle$ is the time-dependent static structure factor and $\Lambda$ is a cutoff 
proportional to the inverse of the length of the surfactant molecule.
It is worth to note that, apart from the presence of renormalized 
coefficients, $\chi^{-1}(k)$ preserves 
the same functional expression of the equilibrium scattering function 
firstly deduced by Teubner and Strey in their seminal paper \cite{TeuStr}.
Such a form yields for $a>0$, $g_0<0$ and $c>0$ a single broad
scattering peak at a non-zero wavevector, and a $k^{-4}$ decay at large $k$ which are properties experimentally observed for a variety of microemulsions
containing comparable amounts of water and oil \cite{Sch,Ceb,Auv,Lich}.
The present Hartree approximation has been used by Huang 
and Muthukumar (see \cite{Mut1} and references therein) for studying 
the steady-state structure factor in block copolymers. 
In the nonequilibrium problem in which we are interested in this paper,
it will be used to study the dynamical process of the 
approach to the steady state. It represents the main simplification 
in our model and it is deemed to be adequate only close to the steady
state (see also \cite{Stin} and references therein, where a similar 
argument was used to study the approach to the steady state of
staggered non-equilibrium particle systems and Ising chains).
Therefore, the linearized Langevin equation for the compositions 
fluctuations is written as
\begin{equation}
\label{eqn_8}
\frac{\partial}{\partial t}\: \varphi(\vec{k},t)= 
\dot{\gamma} \:k_x\: \frac{\partial}{\partial k_y} \varphi(\vec{k},t) - 
 \Gamma \:k^{2}\: \chi (k)\: \varphi(\vec{k},t)
+ \eta(\vec{k},t) 
\end{equation}
The set of Eqs. (\ref{eqn_5a})-(\ref{eqn_8}) is equivalent
to the following Fokker-Planck equation for the probability
distribution of the $\varphi(\vec k)$, $P(\varphi (\vec{k}), t)$ \cite{Parisi}:
\begin{eqnarray}
\label{eqn_9}
&\frac{\partial}{\partial t}\: P( \varphi (\vec{k}), t) =
\int \frac{d \vec{k}}{(2 \pi)^3}\:\{ 
\frac{\delta}{\delta \:\varphi (\vec{k})}
[ - \:\dot{\gamma}\: k_x \frac{\partial\: 
\varphi (\vec{k})}{\partial k_y} + \nonumber\\
&+ \:\Gamma \mid \vec k \mid^2 \left(
\chi (k) \:\varphi (\vec{k})+ 
\frac{\delta}{\delta\: \varphi (- \vec{k})} 
\right) ]
\} \:P(\varphi (\vec{k}), t)  
\end{eqnarray}
Eq.~(\ref{eqn_9}) completely defines the stochastic relaxation process 
in which we are interested and represents the starting point for the 
analysis of the time-dependent static structure factor that will be presented 
in the next session.

\section{Results and Discussion}
\label{Results and Discussion}
\begin{figure}[ht]
\begin{center}
\resizebox{0.40\textwidth}{!}{
\includegraphics{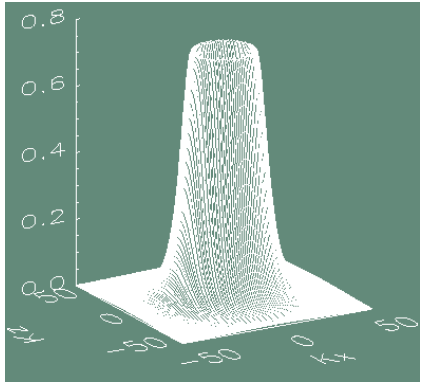}}
\caption{\label{fig1} The structure factor at initial time.}
\end{center}
\end{figure}

Eq.~(\ref{eqn_8}) states a linear dependence of $\varphi(\vec{k},t)$ 
from the noise $\eta(\vec{k},t)$: as the noise is Gaussian distributed, 
then the probability distribution $P [\varphi (\vec k,t)]$ must be Gaussian 
at each time.
In fact the steady solution of Eq.~(\ref{eqn_9}), which represents the
probability distribution for the field $\varphi$ in the steady state, is 
given by 
\begin{equation}
\label{eqn_10} 
P_{0}(\varphi_{\vec{k}})=
\left( \frac{1}{det \hspace{.1cm} C_0(\vec{k}) } \right)
^{1/2}
\exp \left[ -\frac{1}{2} \int \frac{d \vec{k}}{(2 \pi)^3} \hspace{.2cm}
\frac{\varphi (\vec{k})\: \varphi (-\vec{k})}
{C_0 (\vec{k})} \right] \hspace{1cm}
\end{equation}
where $C_0$ stays for the steady-state structure factor 
and has the following expression \cite{Sup1}
\begin{equation}
C_0 (\vec{k}) = 
\int_{0}^{\infty}dt\: e^{- \int_{0}^{t} \omega(k(\zeta))
d\zeta} \: \Gamma \mid \vec k (t)\mid^2
\label{eqn_11}
\end{equation}
where
\begin{equation}
\label{eqn_12}
\vec k (t)=\left( k_x\:,\:k_y\:+\: \frac{\dot{\gamma}\: 
k_x\: t}{2}\:,\:k_z \right)
\end{equation}
and
\begin{equation}
\label{eqn_13}
\omega(k) = \Gamma \:k^2 \:\chi(k)
\end{equation}
is the thermal decaying rate, \textit{i.e.} the rate at which order
parameter fluctuations having a wavelength $\lambda=2 \pi / k$ 
relax at thermal equilibrium (in the absence of flow) \cite{Fred3,OK1}.
Eqs. (\ref{eqn_6}), (\ref{eqn_7}) and (\ref{eqn_11}) form a self-consistent
relationship which accounts for the fluctuations under shear.
By consideration of Eq.~(\ref{eqn_11}) it is apparent that for
$\dot{\gamma}>> \omega(k)$ the composition fluctuations will be
strongly distorted by the flow, \textit{i.e.} the flux forces the 
$C_0(\vec k)$ to be highly anisotropic. The integral in Eq.~(\ref{eqn_11}) is clearly positive, 
as it should be for a variance, and it is possible to verify that 
the solution (\ref{eqn_11}) is unique: in fact every other solution 
different from that does not recover the Ornstein-Zernike form in 
the limit $\dot{\gamma} \rightarrow 0$ \cite{OK1}.
It is worth to note that $C_0 (\vec{k})$ does not depend on the initial 
conditions. It will be shown in the following that the steady-state 
structure factor can be obtained from the time-dependent static one through 
a limit calculation ($t \rightarrow \infty$) after which the correlation 
loses the memory of its initial value.
Because of the Markovian character of the stochastic
process and the translational invariance (see \cite{OK1}), we make
the following ansatz for the time-dependent probability distribution 
of the field $\varphi$ during the dynamical process of the approach 
to steady state, that will be completely proved a posteriori:
\begin{equation}
\label{eqn_14}
P(\varphi (\vec{k}), t)=G(t) \exp \left [- \frac{1}{2}
\int \frac{d \vec{k}}{(2 \pi)^3} \hspace{.2cm} \frac{\varphi (\vec{k})\: 
\varphi( -\vec{k})}{C(\vec{k},t)} \right ]
\end{equation}
where $G(t) = \left[1 / \left( det \hspace {.1cm} C(\vec{k},t)\right)\right]^{1/2}$ 
is a normalization factor, and $C(\vec{k},t)$ is the time-dependent static
structure factor, which is given by the time-dependent solution of 
the following equation  
\begin{equation}
\label{eqn_15}
\frac{\partial C(\vec{k},t)}{\partial t}= \dot{\gamma}\: k_{x}\: 
\frac{\partial  C(\vec{k},t)}{\partial k_y}
-\:2\: \omega(k)\: C(\vec{k},t)\:
+\:2\: \Gamma \:k^2\:
\end{equation}
Eq.~(\ref{eqn_15}) can be obtained formally by taking the time derivative of 
$C(\vec{k},t)=\langle \varphi (\vec{k}, t)\:\varphi (-\vec{k}, t) \rangle$
and using the linearized Langevin Eq.~(\ref{eqn_8}), 
provided a regularization procedure is used for dealing with the difficulties
coming from the fact that time differentiation and averaging do not commute.
Such a regularization entails defining
$\langle \varphi (\vec{k}, t)\: \eta (-\vec{k}, t) \rangle =
\lim_{t' \rightarrow t^{-}}\:\: \langle \varphi (\vec{k}, t)\: \eta (-\vec{k}, t') 
\rangle \:\theta(t-t')$, where the presence of the step function $\theta(t-t')$ 
reflects the causality of the Langevin equation, and then setting this equal 
to $\frac{1}{2}\:\langle \eta (\vec{k}, t)\: \eta (-\vec{k}, t) \rangle$
\cite{Zinn}.\\
The first and second term in the R.H.S. of Eq.~(\ref{eqn_15}) are the convection term and the diffusion term respectively, while the third term is
the fluctuation source. When the shear rate is large enough, the
hydrodynamic force strongly interferes with the thermal decaying
process and a nonlinear effect of suppression of the composition
fluctuations takes place. We shall begin by solving the homogeneous 
problem:
\begin{eqnarray}
\label{eqn_16}
\frac{\partial C(\vec{k},t)}{\partial t}-\dot{\gamma}\: k_{x}\: 
\frac{\partial  C(\vec{k},t)}{\partial k_y} &=&
-\:2\: \omega(k) \:C(\vec{k},t) \\
\label{eqn_17}
C(\vec{k}\:,\:0) &=& C_{init}(k^{2})\:
\end{eqnarray}
where $C_{init}(k^{2})= 1/ \left(a\:+\:g_0\:k^2\:+
\:c\:k^4 \right)$ is the well known isotropic scattering 
function of the microemulsion at initial time, in the absence of 
flow \cite{TeuStr,Gom1}, and its projection on each Cartesian plane is 
that of a circular volcano with radius proportional to the inverse
correlation length of the system, as it is shown in Fig.~(\ref{fig1}).

\begin{figure}[ht]
\begin{center}
\resizebox{0.4\textwidth}{!}{
\includegraphics{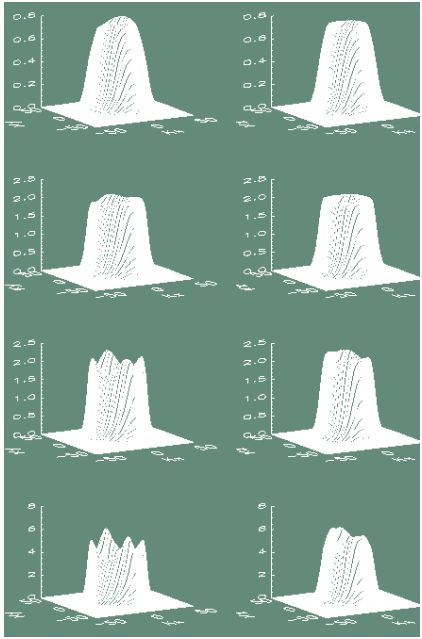}}
\caption{\label{fig2} Projections of the time-dependent static structure factor
on the planes $k_z=0$ (left column) and $k_y=0$ (right column) at
different times t=100, 300, 500, 1000 (from top to bottom). The value of the shear rate is $\dot{\gamma}=2$.}
\end{center}
\end{figure}

\begin{figure}[ht]
\begin{center}
\resizebox{0.4\textwidth}{!}{
\includegraphics{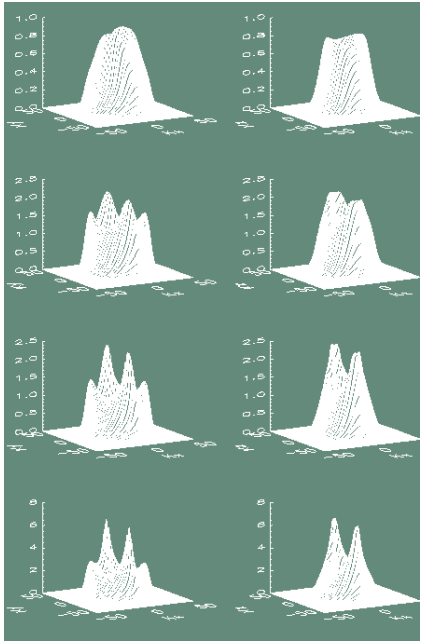}}
\caption{\label{fig3} Projections of the time-dependent static structure factor
on the planes $k_z=0$ (left column) and $k_y=0$ (right column) at
different times t=100, 300, 500, 1000 (from top to bottom). The value of the shear rate is $\dot{\gamma}=8$.}
\end{center}
\end{figure}

The solution of the homogeneous problem is found by using the 
method of characteristics and it has the following form:
\begin{equation}
\label{eqn_18}
C_{hom}(\vec{k},t)= C_{init}(k^{2}(t))
\exp{\left[- \int_{0}^{t} \omega \left(k (t-t')\right)
dt'\right]}
\end{equation}
where $\omega(k)$ has been defined in Eq.~(\ref{eqn_13}).
The full solution of Eq.~(\ref{eqn_15}) with the initial
condition given in Eq.~(\ref{eqn_17}) is 
\begin{equation}
C(\vec{k};t) = C_{hom}(\vec{k};t) + \int_{0}^{t} \Gamma
\mid \vec k (m) \mid^2  
e^{- \int_{0}^{m} \omega(k (\zeta))d \zeta } dm
\label{eqn_19}
\end{equation}
Taking the limit $t \to \infty$, it is straightforward to verify 
that $C(\vec k ,t) \rightarrow C_0 (\vec k)$, \textit{i.e.} the 
time-dependent field distribution $P(\varphi (\vec{k}), t)$ 
converges (in a weak sense) at infinite time towards the steady-state 
distribution $ P_{0}(\varphi (\vec{k}))$: this justifies a posteriori 
the previous ansatz (see Eq.~(\ref{eqn_14})).
Eq.~(\ref{eqn_19}) is the principal result of this paper: it represents
an analytical expression for the time-dependent static structure factor of 
microemulsions in the presence of a plane Couette shear flow.
We shall now consider the behaviour of the structure factor for
short times. In this limit we have
\begin{eqnarray}
\label{eqn_20}
C(\vec{k};t) &=& C_{hom}(\vec{k};0)+ \Gamma\:k^2\:
\left[ 1 - C_{init}(k^{2}(t)) \chi(k) \right] t \:+ \nonumber \\
&+& \frac{d C_{init}(k^2)}{d k^2}\:\dot{\gamma}\:k_x
\:k_y \:t\:+\:O(t^2)
\end{eqnarray}
\textit{i.e.} the scattering intensity grows linearly in t.
A linear dependence on time t has been originally predicted by 
Hennes and Gompper in \cite{Gom4} within a different model. Nevertheless,
the present self-consistent calculation differs from the former
one in two important respects: firstly, the expression given in 
Eq.~(\ref{eqn_20}) shows a dependence on mode-coupling terms 
through $\chi(k)$ that does not appear in \cite{Gom4}: their presence
is likely to be important in predicting the correct approach to the
steady state.  
Secondly, the third term on the R.H.S. of Eq.~(\ref{eqn_20})
explicitly breaks the isotropy of the short time behaviour of 
the scattering function, unless $k_x\:k_y\:=\:0$.
The significance of this anisotropic term is apparent in
the following analysis and is amenable to experimental test.
The predictions of our model are now investigated on the strongly 
structured microemulsion whose steady-state characteristic features 
have already been studied in \cite{Sup1} (to which we also refer for
details on the specific values adopted for the system parameters).
The role of the shear rate $\dot{\gamma}$ as a control parameter
for probing the nonequilibrium properties of microemulsions has been 
already discussed in the Introduction. Therefore, we will now 
use two distinct values of the shear rate ($\dot{\gamma}=2$
and $\dot{\gamma}=8$) which, even if both belonging to the region in which 
shear-thinning behaviour is observed, still produce very different 
mesoscopic structures into the system.
The characteristic morphologies of the time-dependent static structure
factor projections on both the planes $k_z=0$ and $k_y=0$,
are shown in Fig.~(\ref{fig2}) (for $\dot{\gamma}=2$) and in 
Fig.~(\ref{fig3}) (for $\dot{\gamma}=8$) at four different times.   
The first time ($t=100$) is chosen to be sufficiently small to
observe the early deformations of the initial isotropic volcano shape,
while the last one ($t=1000$) is such that the steady-state morphologies
previously found in \cite{Sup1} are recovered.
The effect of the flow on the shape of the structure factor is to
stretch it along a line oriented at $135^o$ to the $k_x$-axis and
rotate it towards the $k_y$-axis ($k_z$-axis) in the $k_z=0$ plane
($k_y=0$ plane). At any fixed time during the relaxation dynamics,
the degree to which the scattering function is stretched and
rotated depends on the value of $\dot{\gamma}$ and grows with it.
It is worth to stress that the short-time morphologies in the 
$k_y=0$ plane are less anisotropic than those in the $k_z=0$ plane
as a result of the absence of the term $\propto \dot{\gamma}\:k_x
\:k_y \:t$ in Eq.~(\ref{eqn_20}) when $k_y=0$: the predicted different degree 
of isotropy should be observable in scattering experiments and 
its possible confirmation would represent an important
test of the present theory.
Finally, the steady-state morphology of the structure factor
strongly depends on $\dot{\gamma}$: in the case $\dot{\gamma}=2$  
the predicted four peaks structure can be related to a biphasic
region with different allignment while above a certain shear rate
the microemulsion is predicted to transform into a lamellar phase
oriented along the flow, as it is observed for $\dot{\gamma}=8$.

\section{Conclusions}
\label{Conclusions}
In the present paper we have studied the response of microemulsion
to the sudden application of a steady plane Couette
shear flow. Analytical expressions for the dynamic scattering
function and the distribution function of composition fluctuations
have been developed and their convergence to steady-state
values proven. The prediction of the model have been investigated 
on a structured microemulsion whose time-dependent static structure factor
has been evaluated and explicitly shown at different times during 
the relaxation dynamics towards the steady state.
The present results can be tested in neutron or X-ray scattering
experiments through which vital informations can be gained to
further developing the analysis of the dynamical process.

\section{Acknowledgments}
\label{Acknowledgments}
The author is grateful to R.~Stinchcombe and L.~Porcar for very 
helpful discussions and encouraging remarks.
He also has a great pleasure in acknowledging the referee for 
making explicit the correct sense in which the name ``structure factor'' has been 
used through the paper.

% \appendix

% \section{}
% \label{}

\end{document}